\documentclass[pss,fleqn,embeddedheads]{w-art}
\usepackage{times}
\usepackage{w-thm}
\usepackage[]{graphicx}
\begin{document}
\DOIsuffix{theDOIsuffix}
\Volume{XX}
\Issue{1}
\Copyrightissue{01}
\Month{01}
\Year{2007}
\pagespan{1}{}
\Receiveddate{\sf 29 September 2006} 
\Accepteddate{\sf 3 December 2006} 
\subjclass[pacs]{75.10.Jm, 03.67.Mn, 71.70.Ej, 73.43.Nq}



\title[Spin-orbital entanglement ]
      {Spin-orbital entanglement near quantum phase transitions }


\author[A. M. Ole\'s]{Andrzej M. Ole\'s\footnote{Corresponding
                      author: e-mail: {\sf a.m.oles@fkf.mpg.de}}\inst{1,2}}
\address[\inst{1}]{Max-Planck-Institut f\"ur Festk\"orperforschung,
                   Heisenbergstrasse 1, D-70569 Stuttgart, Germany}
\address[\inst{2}]{M. Smoluchowski Institute of Physics, Jagellonian 
                   University, Reymonta 4, PL-30059 Krak\'ow, Poland}

\author[P. Horsch]{Peter Horsch\inst{1}}

\author[G. Khaliullin]{Giniyat Khaliullin\inst{1,3}}
\address[\inst{3}]{Kazan Physical-Technical Institute of the 
                   Russian Academy of Sciences, 420029 Kazan, Russia}

\begin{abstract}
Spin-orbital entanglement in the ground state of a one-dimensional 
SU(2)$\otimes$SU(2) spin-orbital model is analyzed using exact 
diagonalization of finite chains. For $S=1/2$ spins and $T=1/2$ 
pseudospins one finds that the quantum entanglement is similar at the 
SU(4) symmetry point and in the spin-orbital valence bond state. 
We also show that quantum transitions in spin-orbital models turn out 
to be continuous under certain circumstances, in constrast to the 
discontinuous transitions in spin models with SU(2) symmetry.\\
{\it [Published in: Phys. Status Solidi (b) {\bf 244}, 2378-2383 (2007).]}
\end{abstract}

\maketitle

Rich magnetic phase diagrams of transition metal oxides and the
existence of quite complex magnetic order with coexisting ferromagnetic 
(FM) and antiferromagnetic (AF) interactions, such as $A$-AF phase in 
LaMnO$_3$ or $C$-AF phase in LaVO$_3$, originate from the intricate 
interplay between spin and orbital degrees of freedom --- 
alternating orbital (AO) order supports FM interactions, whereas ferro 
orbital (FO) order supports AF ones \cite{Goode}. While in many cases 
the spin and orbital dynamics are independent from each other and such 
classical concepts apply, the quantum fluctuations are {\it a priori\/} 
enhanced due to a potential possibility of joint {\it spin-orbital\/} 
fluctuations, particularly in the vicinity of quantum phase transitions 
\cite{Fei97}. Such fluctuations are even much stronger in $t_{2g}$ than 
in $e_g$ systems and may dominate the magnetic and orbital correlations 
\cite{Kha00}, which could then contradict the above classical 
expectations in certain regimes of parameters. Recently it has been 
realized \cite{Ole06} that this novel quantum behavior is accompanied 
by spin-orbital entanglement, similar to that being currently under 
investigation in spin models \cite{Zha06}.  

In general, any spin-orbital superexchange model derived for a 
transition metal compound with a perovskite lattice may be written 
in the following form:
\begin{equation}
\label{som}
{\cal H}=J\sum_{\gamma}\sum_{\langle ij\rangle\parallel\gamma}\left[
    \Big({\vec S}_i\cdot {\vec S}_j+S^2\Big){\hat J}_{ij}^{(\gamma)}
    + {\hat K}_{ij}^{(\gamma)}\right]+{\cal H}_{\rm orb},
\end{equation}
where $\gamma=a,b,c$ labels the cubic axes --- depending on the 
direction of a bond $\langle ij\rangle$ the interactions take a 
different form. The first term in Eq. (\ref{som}) describes the 
superexchange interactions ($J=4t^2/U$ is the superexchange constant, 
where $t$ is the hopping element and $U$ stands for the Coulomb element) 
between transition metal ions in the $d^n$ configuration with spin $S$. 
The orbital operators ${\hat J}_{ij}^{(\gamma)}$ and 
${\hat K}_{ij}^{(\gamma)}$ depend on Hund's exchange parameter
$\eta=J_H/U$, which determines the excitation spectra after a virtual
$d^n_id^n_j\rightarrow d^{n+1}_id^{n-1}_j$ charge excitation. Therefore, 
realistic models of this type (some examples were given recently in 
Ref. \cite{Kha05} and analyzed using the mean field approximation in 
Ref. \cite{Ole05}) are rather involved and contain several terms). 
In addition, orbital interactions can also be induced by the coupling 
to the lattice, and appear in ${\cal H}_{\rm orb}$ term which depends
on a second parameter $V$. 

In the limit of $\eta=0$, however, and for $t_{2g}$ orbitals, 
${\hat J}_{ij}^{(\gamma)}$ and ${\hat K}_{ij}^{(\gamma)}$ operators 
simplify and contain only a scalar product $\vec{T}_i\cdot\vec{T}_j$ 
of $T=1/2$ pseudospin operators which stand for two active $t_{2g}$ 
orbitals along $\gamma$, so the SU(2) symmetry 
follows both for spin and for pseudospin interactions. Here we shall 
discuss primarily a one-dimensional (1D) model in this idealized 
situation and investigate the dependence of spin-orbital entanglement 
on the type of underlying interactions. To characterize the ground 
state we evaluated intersite spin, orbital and composite spin-orbital
correlations, defined as follows for a bond $\langle ij\rangle$  
\cite{Ole06}:
\begin{eqnarray}
\label{st}
S_{ij}&=&\langle{\vec S}_i\cdot {\vec S}_j\rangle/(2S)^2, \hskip 2.4cm
T_{ij} = \langle{\vec T}_i\cdot {\vec T}_j\rangle, \\
\label{cij}
C_{ij}&=&
\big[\big\langle({\vec S}_i\cdot{\vec S}_j)
                ({\vec T}_i\cdot{\vec T}_j)\big\rangle
     -\big\langle{\vec S}_i\cdot{\vec S}_j\big\rangle
      \big\langle{\vec T}_i\cdot{\vec T}_j\big\rangle\big]/(2S)^2.
\end{eqnarray}
Note that $C_{ij}$ quantifies the quantum entanglement --- if $C_{ij}<0$ 
spin and orbital operators are entangled and mean field approximation,
i.e., decoupling of spin and pseudospin operators is not justified. 

\begin{SCfigure}[1][t!]
\includegraphics[width=.54\textwidth]{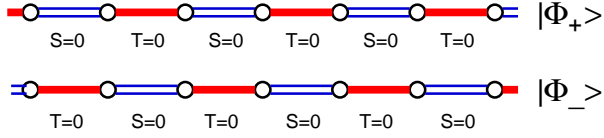}
\caption
{
Two degenerate SOVB ground states $|\Phi_{\pm}\rangle$ with alternating 
spin ($S=0$) and orbital ($T=0$) singlets for the SU(2)$\times$SU(2) 
model (\ref{su2xsu2}) at the dimer point $p=3/4$. 
} 
\label{fig:sovb}
\end{SCfigure}

The 1D SU(2)$\times$SU(2) model,
\begin{equation}
\label{su2xsu2gen}
{\cal H}_J=J\sum_{i}
    \Big({\vec S}_i\cdot {\vec S}_{i+1}+x\Big)
    \Big({\vec T}_i\cdot {\vec T}_{i+1}+y\Big),
\end{equation}
has two parameters $x$ and $y$. Its phase diagram in the $(x,y)$ plane 
consists of five 
distinct phases which result from the competition between effective 
AF and FM spin (AO and FO pseudospin) interactions on the bonds 
\cite{Aff00}. First of all, the spin and pseudospin correlations are 
FM-FO, $S_{ij}=T_{ij}={1\over 4}$, if $x<-{1\over 4}$ 
and $y<-{1\over 4}$. Then the ground state is characterized by the 
maximal values of both total quantum numbers, ${\cal S}={\cal T}=N/2$, 
where $N$ is the chain length, its degeneracy is $d=(N+1)^2$, and the 
quantum fluctuations are suppressed. Two other phases, with either spin 
FM or pseudospin FO correlations, show also no entanglement as the wave 
function factorizes, i.e., spin and orbital operators may be then still 
decoupled from each other. 
In these cases only either spin or pseudospin quantum fluctuations
occur. On the contrary, in the other two phases both spin and orbital
correlations are negative \cite{Li05}, so one expects that quantum 
entanglement takes over.

\begin{SCfigure}[2][b!]
\includegraphics[width=.66\textwidth]{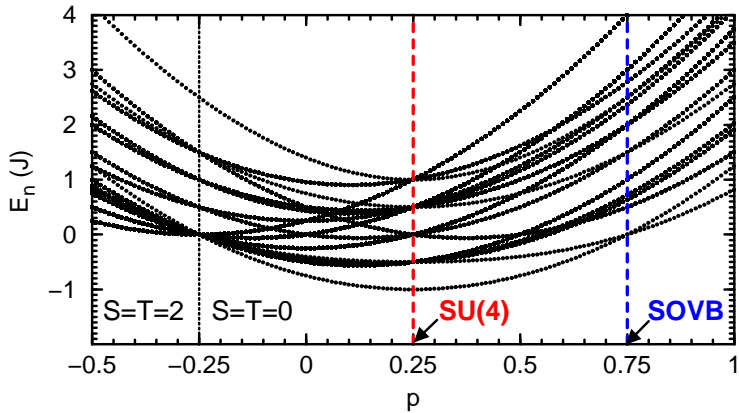}
\caption
{
Energy spectrum for the $S=1/2$ spin-orbital chain (\ref{su2xsu2})
of $N=4$ sites for increasing $p$. Some of the 256 eigenenergies have 
high degeneracy $d$. At $p=-0.25$ the ground state 
changes from high spin-orbital state (${\cal S}={\cal T}=2$, $d=25$)  
to spin-orbital singlet state (${\cal S}={\cal T}=0$, $d=1$); 
its degeneracy at $p=-0.25$ is $d=194$. 
The SU(4) point ($p=0.25$, $d=1$) and the SOVB point ($p=0.75$, $d=2$) 
are marked by vertical dashed lines.  
} 
\label{fig:dos}
\end{SCfigure}

This motivates our study of the SU(2)$\otimes$SU(2) model along the line 
$p=x=y$, where $S=1/2$ spins and $T=1/2$ pseudospins appear on equal 
footing. However, in order to allow also for a larger value of spin $S$ 
per site, we write the Hamiltonian along this line in the following way, 
\begin{equation}
\label{su2xsu2}
{\cal H}_J=J\sum_{i}
    \Big({\vec S}_i\cdot {\vec S}_{i+1}+{4\over 3}pS(S+1)\Big)
    \Big({\vec T}_i\cdot {\vec T}_{i+1}+p\Big),
\end{equation}
with ${4\over 3}pS(S+1)=p$ for $S=1/2$. Interestingly, for 
$p={3\over 4}$ the model given by Eq. (\ref{su2xsu2}) has an exact 
doubly degenerate ground state $|\Phi_{\pm}\rangle$, with alternating 
spin and orbital singlets forming a spin-orbital valence bond (SOVB) 
phase (Fig. 1). These states resemble the Majumdar-Ghosh valence bond 
states in the 1D spin model with next-nearest interactions \cite{Maj69} 
and have exact energy $E_0=0$ given by either spin or orbital singlet. 
Each $|\Phi_{\pm}\rangle$ state is a matrix product state in both spin 
and orbital sector \cite{Kol98}. For $S=1/2$ and $p={1\over 4}$ one 
recovers the celebrated SU(4) model, with all three correlation 
functions: $S_{ij}$, $T_{ij}$, and 
${4\over 3}\langle ({\vec S}_i\cdot{\vec S}_{i+1})
                   ({\vec T}_i\cdot{\vec T}_{i+1})\rangle$ being equal 
to each other \cite{Fri99}. The energy spectra in these two cases look 
quite differently for an $N=4$ site chain --- an equidistant energy 
spectrum with high degeneracies of each excited state with energies 
$(-1+{n\over 2})J$ where $n=0,1,\cdots,4$ is found at the 
SU(4) point, while the spectrum consists of many eigenenergies with 
lower degeneracies at the SOVB point 
(Fig. 2). It is {\it a priori\/} not clear in which of 
these two points the quantum entanglement is stronger, but one might 
expect that it would be more pronounced at the SU(4) point. 

In fact, it was shown recently that a reduced von Neumann entropy is 
there maximal \cite{Che06}. While the spin-orbital entropy could help 
to understand the consequences of the entanglement at finite temperature 
which might trigger phase transistions to phases with strong dimer 
correlations \cite{Hor03} observed in experiment \cite{Ulr03}, here we 
suggest that a useful measure of {\it entanglement in the ground state\/} 
is the intersite spin-orbital correlation function (\ref{cij}) which 
quantifies the error of the mean field decoupling of spin and orbital 
operators on individual bonds. The intersite spin, orbital and 
spin-orbital correlations demonstrate a quantum phase transition between 
the high spin-orbital FM-FO state (${\cal S}={\cal T}=2$) and the 
singlet entangled state (${\cal S}={\cal T}=0$) at $p=-{1\over 4}$ in 
the 1D model (\ref{su2xsu2}) for $N=4$ and $N=8$ sites, see Fig. 3. 
In fact, the Hamiltonian (\ref{su2xsu2}) is then a product of singlet 
projection operators in spin and orbital space, so the FM-FO state has 
the lowest possible energy $E_0=0$ and is degenerate with several other
states (Fig. 2). 
 
\begin{figure}[t!]
\includegraphics[width=11cm]{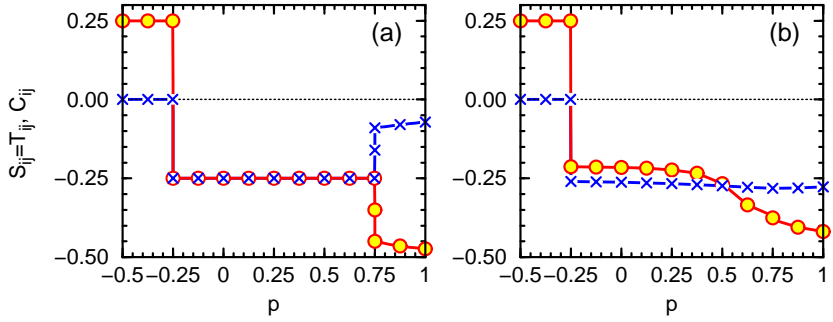}
\caption
{
Intersite spin and orbital correlations $S_{ij}=T_{ij}$ (\ref{st}) 
(circles), and composite spin-orbital correlations $C_{ij}$ (\ref{cij}) 
(crosses), as obtained for the 1D spin-orbital chains for increasing 
$p$ in Eq. (\ref{su2xsu2}), with:
(a) $N=4$, and
(b) $N=8$ sites.
} 
\label{fig:corr}
\end{figure}

As expected, $C_{ij}=0$ for $p<-{1\over 4}$, and the spin and orbital 
degrees of freedom disentangle. The situation is qualitatively 
different above the quantum transition at $p={1\over 4}$, where 
$C_{ij}\simeq -0.25$ ($C_{ij}=-0.2595$, $-0.25$ for an $N=8$, 4 site 
chain at $p=-0.249$, respectively). Although we are not interested here 
in accurate quantitative values of the respective intersite 
correlations, we observe that the $N=8$ site chain comes already quite 
close to the thermodynamic limit at the SU(4) point --- 
we found $S_{ij}=T_{ij}=-0.223$ in place of $-0.215$ for a chain of 
$N=100$ sites (which almost reproduces the Bethe ansatz result for an
infinite chain) \cite{Fri99}. Furthermore, in the case of $N=8$ site 
chain, the data show a smooth decrease of $S_{ij}=T_{ij}$ correlations 
with increasing $p$, indicating the tendency towards AF-AO spin-orbital 
correlations in the regime of large $p$. This is markedly different 
from the $N=4$ site chain, where these correlations are constant 
($S_{ij}=T_{ij}=-0.25$) in the entire range of 
$-{1\over 4}<p<{3\over 4}$, i.e., up to the SOVB point, where the data 
suggest a second quantum phase transition to the AF-AO state (Fig. 3a). 
In fact, this transition is a finite size effect and is replaced by a 
{\it smooth crossover\/} towards smaller values of both $S_{ij}$ and 
$T_{ij}$ correlations in an $N=8$ site chain (Fig. 3b).    

\begin{figure}[t!]
\includegraphics[width=11cm]{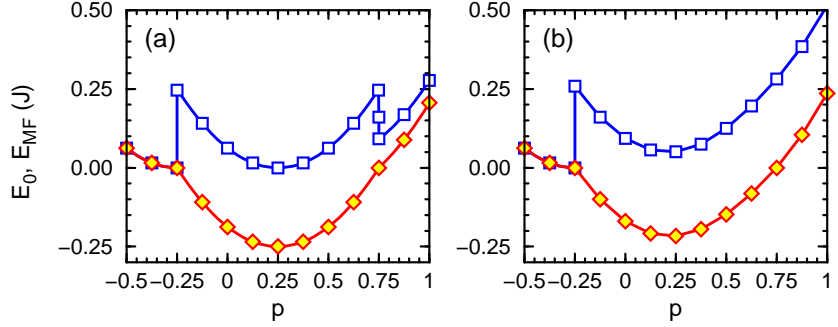}
\caption
{
Ground state energy $E_0$ (diamonds), and mean field energy $E_{\rm MF}$ 
(squares) per bond, as obtained for the 1D spin-orbital chains for 
increasing $p$ in Eq. (\ref{su2xsu2}), with:
(a) $N=4$, and
(b) $N=8$ sites. 
} 
\label{fig:ene}
\end{figure}

It is quite remarkable that the composite spin-orbital correlation 
function (\ref{cij}) has a shallow minimum $C_{ij}=-0.2812$ for an 
$N=8$ site chain at the SOVB point, in spite of decreased individual 
spin and orbital correlations $S_{ij}=T_{ij}=-0.3749$, suggesting 
rather independent spin and orbital dynamics. Therefore,  
precisely at this point the entanglement in the exact wave functions 
$|\Phi_{\pm}\rangle$ shown in Fig. 1 is larger than at the SU(4) 
symmetric point, where $C_{ij}=-0.2667$. We emphasize that sufficiently 
long chain with $N=8$ sites was necessary to reach this conclusion, 
while shorther chains give inconclusive results, either due to 
particular stability of the SU(4) singlet for $N=4$ sites, 
or due to frustrated four-site singlet correlations for $N=6$ sites 
which result in reduced values of $|C_{ij}|$ and $E_{\rm MF}$ (Table I). 

\begin{vchtable}[b!]
\vchcaption{
Intersite spin correlations $S_{ij}$,
       orbital correlations $T_{ij}$ (\ref{st}), and composite
  spin-orbital correlations $C_{ij}$ (\ref{cij}),
as well as mean field $E_{\rm MF}$ energy per site (\ref{emf}),
obtained for various SOVB phases: 
for $S=1/2$ with increasing cluster size of $N=4$, 6 and 8 sites,
and for $S=1$ and $S=3/2$ with clusters of $N=4$ sites. 
By construction, the exact ground state energy is $E_0=0$ in all cases.
}
\vskip .1cm
\begin{tabular}{cccccccc} \hline\hline
 $S$ & $N$ & &  $S_{ij}$ &  $T_{ij}$ &  $C_{ij}$ & & $E_{\rm MF}$ ($J$)\cr
\hline
 1/2 &  4  & & $-0.3500$ & $-0.3500$ & $-0.1600$ & & 0.1600  \cr
     &  6  & & $-0.3735$ & $-0.3735$ & $-0.1417$ & & 0.1417  \cr
     &  8  & & $-0.3749$ & $-0.3749$ & $-0.2812$ & & 0.2812  \cr
\hline
  1  &  4  & & $-0.2429$ & $-0.3643$ & $-0.0992$ & & 0.0992  \cr
 3/2 &  4  & & $-0.2050$ & $-0.3690$ & $-0.0806$ & & 0.0806  \cr
\hline
\end{tabular}
\label{tab:ex}
\end{vchtable}

In the entire regime of singlet states (${\cal S}={\cal T}=0$) in the 
1D spin-orbital model Eq. (\ref{su2xsu2}), one finds large 
corrections to the mean field energy normalized per one bond,
\begin{equation}
\label{emf}
E_{\rm MF}=
    \Big(\Big\langle{\vec S}_i\cdot{\vec S}_{i+1}\Big\rangle+p\Big)
    \Big(\Big\langle{\vec T}_i\cdot{\vec T}_{i+1}\Big\rangle+p\Big),
\end{equation}
as shown in Fig. 4. This demonstrates that one should not decouple 
spin and orbital operators from each other as this procedure suppresses 
an essential part of joint spin-orbital quantum fluctuations and leads 
to inconclusive results. For the same reason, effective magnetic 
exchange constants obtained by averaging over orbital operators 
do not represent a useful concept and cannot be introduced 
in the entangled regime, similar to the realistic spin-orbital models 
for transition metal oxides \cite{Kha05,Ole05}.

We have verified that quantum entanglement is significant in the SOVB 
wave functions (Fig. 1) also at higher values of spin $S$ in Eq.
(\ref{su2xsu2}). However, when spins increase, the AF correlations 
weaken and AO correlations are enhanced. This may be seen as partial 
decoupling of slow and fast quantum fluctuations associated with large 
(small) spin (pseudospin) value, respectively. Indeed, the values of
$|C_{ij}|$ and of $E_{\rm MF}$ decrease with increasing $S$, as shown 
in Table I.   

\begin{figure}[t!]
\includegraphics[width=11cm]{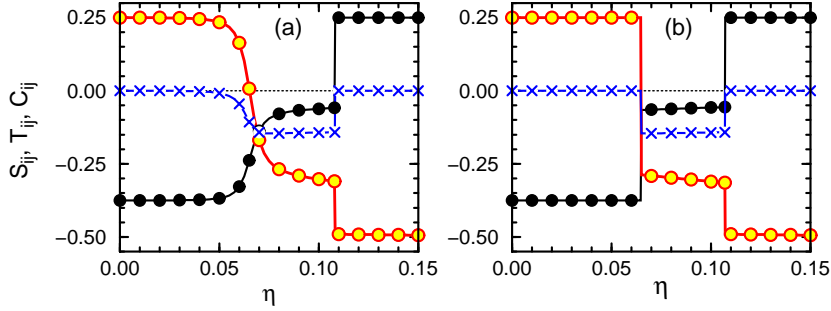}
\caption{
Intersite spin $S_{ij}$ (filled circles),
orbital $T_{ij}$ (empty circles), and composite 
spin-orbital $C_{ij}$ (crosses) correlations, as obtained with an $N=4$ 
site chain for increasing Hund's exchange $\eta$ in:
(a) the vanadate model (\ref{som}) of Ref. \cite{Hor03}, and for
(b) the same model without orbital fluctuating terms in Eq. (\ref{som}).
Orbital interactions resulting from GdFeO$_3$-type distortions are given 
by ${\cal H}_{\rm orb}=-V\sum_iT_i^zT_{i+1}^z$ with $V=J$, and favor FO
order at small $\eta$.
} 
\label{fig:d2}
\end{figure}

Finally, we would like to emphasize that the quantum entanglement is
also a common feature of spin-orbital models (1) derived for transition 
metal oxides when the parameter $\eta=J_H/U$ is small \cite{Ole06}. 
(Note that large $\eta$ plays here a similar role to small 
$p$ in the SU(2)$\otimes$SU(2) spin-orbital model.) Also in such cases 
mean field procedure fails and joint spin-orbital fluctuations $C_{ij}$
dominate in the ground state. Such models exhibit even 
richer behavior than the idealized SU(2)$\otimes$SU(2) spin-orbital 
model considered above. In fact, the SU(2) symmetry concerns only spin 
interactions while it is removed in the orbital sector due to the 
analytic structure of Coulomb interactions and the multiplet structure 
of excited states at $\eta>0$. As a result, one may find 
{\it continuous orbital transitions\/} of the crossover type even when 
the orbital quantum number ${\cal T}$ changes, as shown in Fig. 5a for 
the vanadate model of Ref. \cite{Hor03} with $S=1$ spins. However, when 
the respective interactions are simplified to classical Ising terms 
(Fig. 5b), such transitions appear to be first order \cite{Kaw04}, and 
are qualitatively similar to those encountered usually in the spin 
sector.

Summarizing, we have demonstrated large quantum entanglement in the  
SU(2)$\otimes$SU(2) spin-orbital model for $S=T=1/2$, which appears to
be somewhat stronger in the exact spin-orbital valence bond states of
Fig. 1 than at the SU(4) symmetric point. While the spin-orbital 
entanglement occurs in a discontinuous way in quantum spin transitions, 
it may appear gradually in continuous quantum transitions which involve 
orbital degrees of freedom.

\begin{acknowledgement}
This work was supported by the Polish Ministry of Science and Education
Project No. 1~P03B~068~26.
\end{acknowledgement}

\end{document}